\documentclass[journal,comsoc]{IEEEtran}
\usepackage[T1]{fontenc}
\usepackage{graphicx}
\usepackage{comment}
\usepackage{mathtools}

\usepackage{tikz}
\newcommand*\circled[1]{\tikz[baseline=(char.base)]{
            \node[shape=circle,draw,inner sep=.2pt] (char) {#1};}}
\usepackage[pagebackref=true,breaklinks=true,colorlinks,bookmarks=false]{hyperref}
\usepackage{color, colortbl}
\definecolor{Gray}{gray}{0.9}
\definecolor{LightCyan}{rgb}{0.88,1,1}

\ifCLASSINFOpdf
\else
 
\fi
\usepackage{amsmath}
\usepackage[cmintegrals]{newtxmath}
\begin{document}
\title{PTSR: Patch Translator for Image Super-Resolution}

\author{Neeraj Baghel, Shiv Ram Dubey, Satish Kumar Singh
        
\thanks{Neeraj Baghel, Shiv Ram Dubey and Satish Kumar Singh are with the Computer Vision and Biometrics Lab at Department of Information Technology, Indian Institute of Information Technology Allahabad, Prayagraj, India (email: neerajbaghel@ieee.org, srdubey@iiita.ac.in, sk.singh@iiita.ac.in). 
}
}

\markboth{PTSR}%
{Baghel \MakeLowercase{\textit{et al.}}}

\maketitle

\begin{abstract}
Image super-resolution generation aims to generate a high-resolution image from its low-resolution image. However, more complex neural networks bring high computational costs and memory storage. It is still an active area for offering the promise of overcoming resolution limitations in many applications. In recent years, transformers have made significant progress in computer vision tasks as their robust self-attention mechanism. However, recent works on the transformer for image super-resolution also contain convolution operations. We propose a patch translator for image super-resolution (PTSR) to address this problem. The proposed PTSR is a transformer-based GAN network with no convolution operation. We introduce a novel patch translator module for regenerating the improved patches utilising multi-head attention, which is further utilised by the generator to generate the $2\times$ and $4\times$ super-resolution images. The experiments are performed using benchmark datasets, including DIV2K, Set5, Set14, and BSD100. The results of the proposed model is improved on an average for $4\times$ super-resolution by 21.66\% in PNSR score and 11.59\% in SSIM score, as compared to the best competitive models. We also analyse the proposed loss and saliency map to show the effectiveness of the proposed method. 
\end{abstract}

\IEEEpeerreviewmaketitle

\section{Introduction}
\IEEEPARstart{I}{mage} super-resolution is an important computer vision task, which refers to reconstructing the corresponding high-resolution image from the given low-resolution counterpart \cite{L_unsupervised} \cite{vt_single} \cite{vt_fast} \cite{vt_robust}. The single image super-resolution (SISR) methods have been widely used in advanced visual tasks, such as compression \cite{L_csr}, facial analysis \cite{EG_faceSR} \cite{vt_modeling}, video super-resolution \cite{EG_videoSR}, Stereoscopic Image \cite{vt_deep} satellite and aerial imaging \cite{EG_satelliteSR}, security and surveillance imaging \cite{EG_SurveillanceSR}, and many more.

In recent years, image super-resolution has achieved great recognition \cite{survey2020IeeeTransPatten} and various deep learning approaches have been proposed to address super-resolution problems, such as convolution neural networks (CNNs) \cite{SRCNNr1,EDSR-baseline,CARN,IGNN,OISR-RK3,IMDN}, generative adversarial networks (GANs) \cite{SRGAN, esrgan, DUSGAN, GMGAN} and transformer networks \cite{SAN, RCAN, ESRT, IPT}.

\paragraph{Image Super-resolution using CNNs}
A CNN model consisting of three convolution layers is used in Super-Resolution Convolutional Neural Network (SRCNN) \cite{SRCNNr1}. FSRCNN \cite{FSRCNNr2} proposes a post-upsampling mode to reduce the computational cost. Enhanced Deep Residual Network (EDSR) \cite{EDSR-baseline} and Cascading Residual Network (CARN) \cite{CARN} use residual blocks. VDSR \cite{VDSRr3} uses deep CNN for image super-resolution. IMDN \cite{IMDN} proposes an efficient and lightweight CNN model for faster image super-resolution. Internal Graph Neural Network (IGNN) \cite{IGNN} exploits image's cross-scale patch recurrence property. Resolution-Aware Network
for Image Super-Resolution \cite{vt_raisr}, MIPN \cite{L_MIPN} and AMNet \cite{l_AMNET} exploit asynchronous multi-scale network for image super-resolution.  Though these methods utilize different CNN models and characteristics, their performance is limited due to the lack of the proper utilization of global context. Moreover, these models cannot focus on the important regions requiring more attention, such as high-frequency and blur regions.

\begin{figure}[t]
\centering
\includegraphics[width=1.06\linewidth]{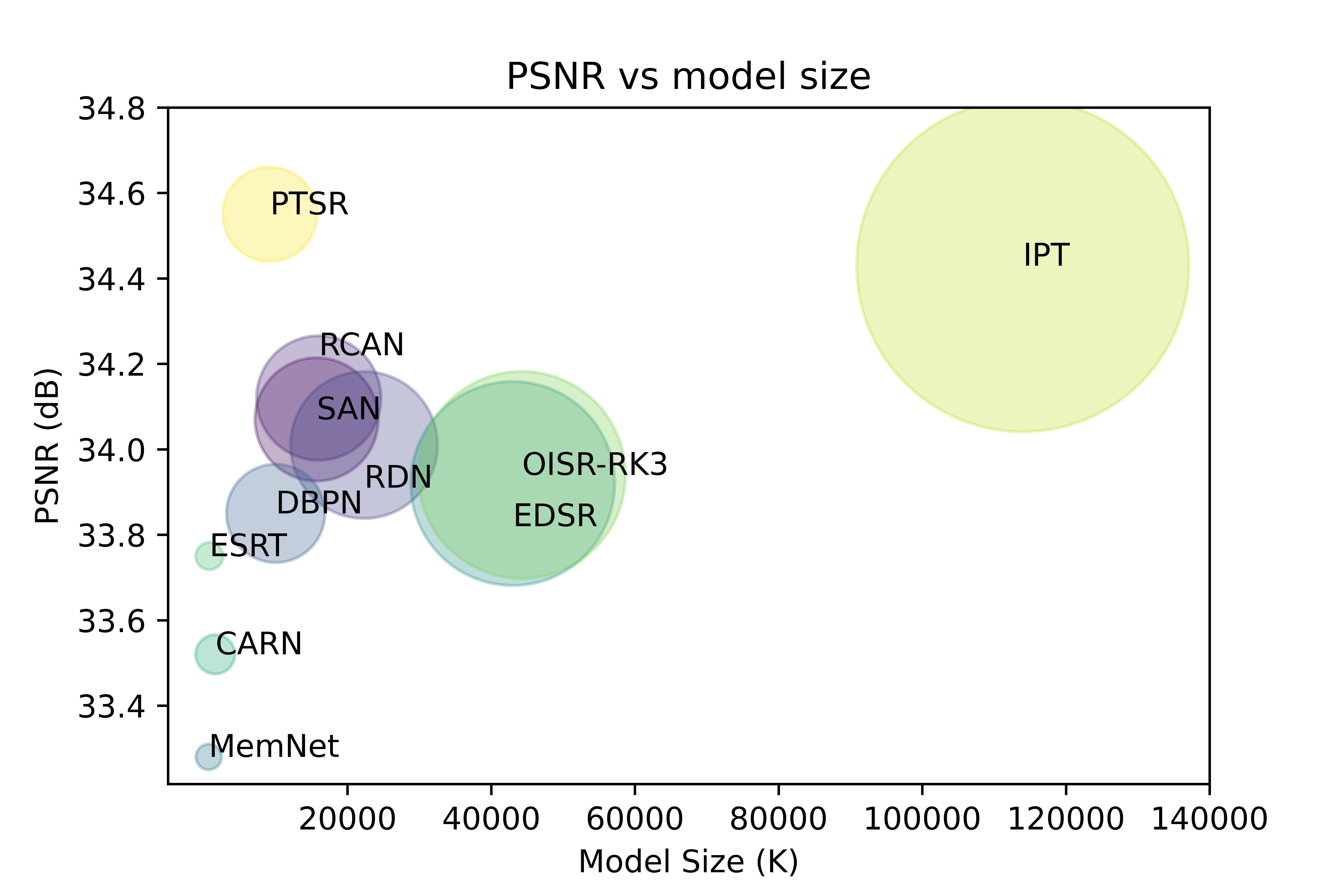}
\caption{Average quantitative performance in terms of PSNR vs model size. We plot results of the proposed PTSR and recently proposed state-of-the-art deep learning models for image super-resolution, including IPT \cite{IPT}, SAN \cite{SAN}, RCAN \cite{RCAN}, RDN \cite{RDN}, DBPN \cite{DBPN}, EDSR \cite{EDSR-baseline}, OISR-RK3 \cite{OISR-RK3}, ESRT \cite{ESRT}, CARN \cite{CARN}, and MemNet \cite{Memnet}. The proposed PTSR method achieves a high performance in-spite of being a very efficient model.}
\label{fig:PSNRvsModel}
\end{figure}

\begin{figure*}[!tbh]
\centering
\includegraphics[width=\linewidth]{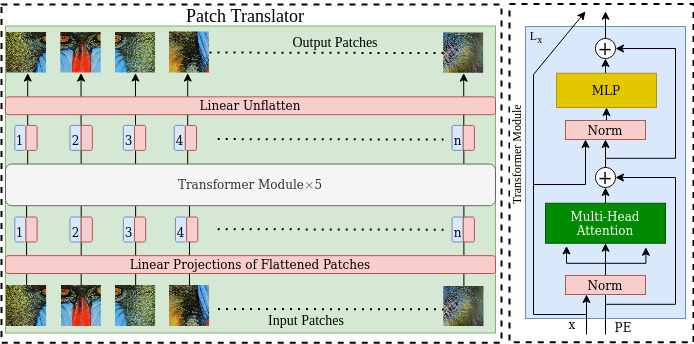}
\caption{Proposed convolution-free patch translator for image translation based on multi-head attention driven transformer. It divides image $I$ into vector patches $V_i$ with positional embedding $PE$. Then the proposed transformer module is used to convert it back into vector patches to generate the image.}
\label{fig:PatchTranslator}
\end{figure*}

\paragraph{Image Super-resolution using GANs}
Generative Adversarial Networks (GANs) contain a generator to transform the image and a discriminator to distinguish between the actual target image and the transformed image \cite{goodfellow2014generative}, \cite{PIX}. The image super-resolution task has also witnessed enormous success using GAN-based approaches. SRGAN \cite{SRGAN} and ESRGAN \cite{esrgan} utilize a CNN-based generator and discriminator networks for image super-resolution. ZoomGAN \cite{L_ZoomGAN} exploits the residual dense blocks and focuses on a particular context. GMGAN \cite{GMGAN} and DUS-GAN \cite{DUSGAN} improve the perceptual quality with GMSD and QA quality losses, respectively. Though GAN-based models have shown promising performance, their generator and discriminator networks miss to utilize the global context effectively leading to limited learning capability.

\paragraph{Image Super-resolution using Transformers}
The key idea of the transformer is ``self-attention” \cite{attention2017vaswani}, which helps to capture the long-term information between sequence elements leading to better utilization of global context. The vision transformer has been very successful for different applications, including image classification, image retrieval, etc. \cite{vit}, \cite{dubey2022vision}. However, the working of vision transformer is similar to the transformer, except the conversion of image patches into embeddings. Few attempts have also been made to utilize the transformers for image super-resolution. Second-order attention network (SAN) \cite{SAN}, Residual channel attention network (RCAN) \cite{RCAN} and meta-attention \cite{L_meta_attention} utilize the residual block and attention module for image super-resolution. Efficient transformer for single image super-resolution (ESRT) \cite{ESRT} uses the transformer with CNN structure, and IPT \cite{IPT} exploits a pre-trained model with the ImageNet benchmark. These transformer-based models for image super-resolution are not convolution-free. They are not used in the GAN framework and hence miss the generative power in the high-resolution image synthesization. However, the recent progress in transformer networks \cite{dubey2023transformer}, such as ViTGAN \cite{vitgan} and TransGAN \cite{Transgan}, indicates the improved performance of transformers in the GAN framework. 

Motivated by transformer-based GAN's success, we propose a Patch Translator for Image Super-Resolution (PTSR). Following are the contributions of this paper:
\begin{itemize}
    \item We propose a convolution-free transformer-based network for both generator and discriminator network. The proposed PTSR generator framework produces $2\times$ images by utilize the patch translator module as a backbone. 
    \item As the primary transformer is not suitable at the patch level, we introduce the patch translator module based on a vision transformer which can be used for any image-to-image translation task. Hence, the proposed model retains the local context through the patch processing and global context through the transformer module. 
    \item The proposed transformer module contains the positional embedding and image information separately for learning the distribution while preserving the patch location. It is beneficial for image-to-image translation tasks.
    \item We conduct extensive experiments which show that our established model is memory \& computation efficient and observe superior performance using PTSR model as compared to SOTA (see Fig. \ref{fig:PSNRvsModel}).
    
\end{itemize}

\begin{figure*}[t]
\centering
\includegraphics[width=\linewidth]{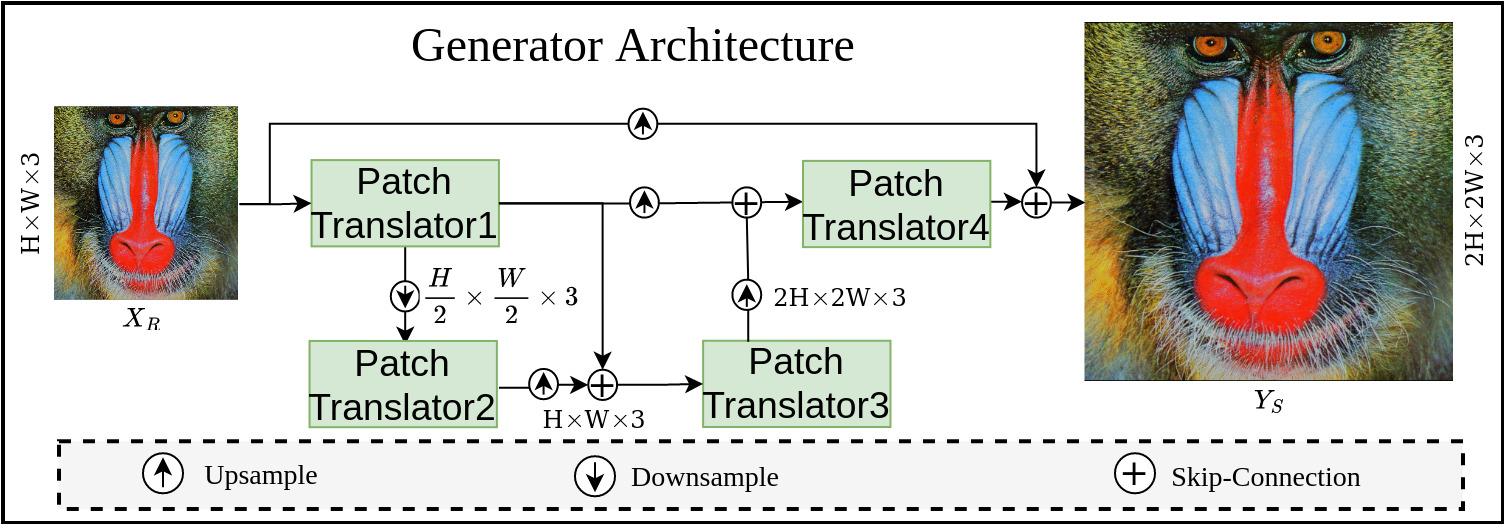}
\caption{The proposed PTSR generator framework $G_{R2S}$ uses a patch translator for image super-resolution. It takes $X_R$ and learns the features (i.e., the difference between $\uparrow X_R$ and $Y_R$). }
\label{fig:Generator}
\end{figure*}

\section{Proposed Patch Translator for Image Super-Resolution}
This section describes the generator and discriminator's overall structure for super-resolution. We introduce a patch translator for regenerating the improved patches utilising multi-head attention with a vision transformer. 
First, Section \ref{Patch Translator} describes our proposed patch translator architecture. Section \ref{Patch Translator based Generator} introduces the proposed Generator architecture in detail. Then, Section \ref{Transformer based Discriminator} introduces the proposed Discriminator architecture overall in detail.

\subsection{Patch Translator} \label{Patch Translator}
The patch translator based on transformer modules exploits the global relationship better than the local one, essential to synthesising the images at high resolution. The proposed Patch Translator is illustrated in Fig. \ref{fig:PatchTranslator}.

\paragraph{Patch Module with Embedding} 
In this module, any given image $I \in \mathbb{R}^{m,m,3}$ is divided into `n' non-overlapping image patches $I_i \in \mathbb{R}^{k,k,3}$, where $i = 1,2,3,... n$ and $k$ is the patch size. For non-overlapping patches stride length, $S_l$ is equivalent to patch size $k$; here, $k=8$. We do not consider the overlapping image patches in the generator module to reduce the number of parameters. The image patches $I_i$ are reshaped into vectors $V_i \in \mathbb{R}^{1,d}$, where $d=k\times k\times 3$. We introduce the positional embedding ($PE$), which consists of dimension $n$ for each patch having a random vector $RV$ of dimension $d$ using a linear projection with parameter $W_{PE}$ as follows,
\begin{equation}
PE_i = RV_i \times W_{PE}
\end{equation}
where $RV_i \in \mathbb{R}^{1,d}$ is the flattened vector corresponding to the $i^{th}$ patch, $W_{PE} \in \mathbb{R}^{d,d}$ is the parameter matrix and $PE_i \in \mathbb{R}^{1,d}$ is the embedding w.r.t. $i^{th}$ patch. This $PE$ is forwarded to the transformer module and corresponding image patches.

\paragraph{Transformer Module}
The transformer module has an $L$ stack of transformer blocks to increase the learning capacity. The positional embedding $PE \in \mathbb{R}^{n,d}$ and vectors $V_i \in \mathbb{R}^{1,d}$ are given as input to the transformer and transformed into an output of embedding having the same dimension. In the transformer block $L_x$, first linear normalization using Self-modulated LayerNorm \cite{vitgan} takes vector patches $V_i$ and their embedding $PE$ as input. It gives the normalized output as $l_i \in \mathbb{R}^{n,d}$ and is forwarded as input to multi-head attention by generating three parametric projections as Query (Q), Key (K), and Value (V). The multi-head attention produces the output as self-attention features $F_S$ by utilizing the $V$ Value vectors and attention weights $W_A$ generated from $Q$ Query and $K$ Key vectors. Then the residual connection is used as,
\begin{equation}
F_R = PE + F_S.
\end{equation}
The output of residual connection $F_R$  with image patches is normalized with Self-modulated LayerNorm. A multi-layer perceptron module is applied on the output of normalization with a linear projection to $\text{mlp}_{dim} = d \times \text{mlp}_{ratio}$ dimension, GELU activation and a linear projection back to $d$ dimension.

\begin{figure*}[t]
\centering
\includegraphics[width=\linewidth]{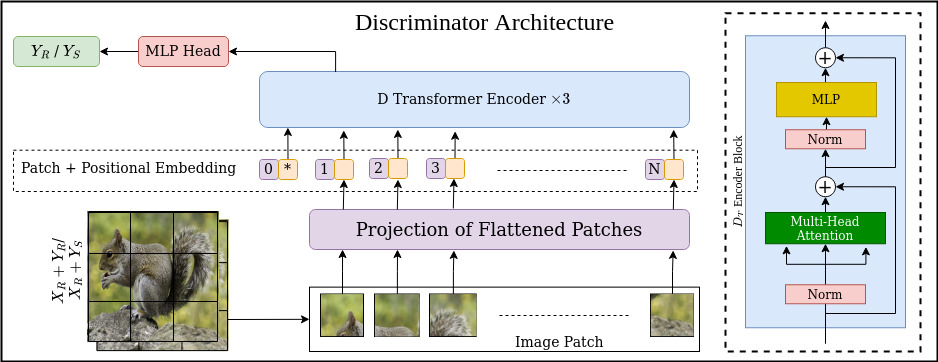}
\caption{The Vision Transformer based Discriminator Network $D_T$ for image super-resolution. It takes 
$X_R+Y_R$ (concatenated low-resolution with High-resolution) or $X_R+Y_S$ (concatenated low-resolution with Super-resolution) as input. It learns difference between $Y_R$ (High-resolution) and $Y_S$ (Super-resolution) while having input as $X_R$ (Low-resolution). }
\label{fig:modeld}
\end{figure*}

\subsection{Patch Translator based Generator}\label{Patch Translator based Generator}
A generator network in image super-resolution takes a low-resolution $X_R$ image as input and produces the corresponding super-resolution $Y_S$ image as output. The proposed generator utilises a patch translator mechanism referred to as $G_{R2S}$. It does not contain any convolution operation. The proposed $G_{R2S}$ network is designed to progressively generate the super-resolution images at higher scales, first by down-sampling and then up-sampling using the patch translator as depicted in Fig. \ref{fig:Generator}. In this, the low-resolution image $X_R \in \mathbb{R}^{H, W,3}$  is given as input to the $G_{R2S}$ network, where $(H$,$W$,$3)$ represents the height, width and color channels of the $X_R$ input image. This input image $X_R$ pass through a patch translator structure referred as $PT_1$ and transform the $X_R \in \mathbb{R}^{H,W,3}$ into $X_{F1} \in \mathbb{R}^{H,W,3}$. Then another patch translator $PT_2$ process this down-sampled feature $\downarrow$ $X_{F1} \in \mathbb{R}^{H/2,W/2,3}$ to $X_{F2} \in \mathbb{R}^{H/2,W/2,3}$. It helps the network learn the important characteristics of down-sampled space in a super-resolution context. Then $X_{F2}$ is up-sampled $\in \mathbb{R}^{H,W,3}$ and added $X_{F1}$ with skip-connection. This combined information is passed to patch translator $PT_3$ to learn the vector relationship between $X_{F1}$ and $X_{F2}$ which have the features at different scales and produce features as $X_{F3} \in \mathbb{R}^{H, W,3}$. Then $X_{F3}$ is up-sampled $\in \mathbb{R}^{2H,2W,3}$ and added with up-sampled $\uparrow$ $X_R$ with skip-connection. This combined information $\uparrow$ $X_{F1}$ and $X_{F3}$ is passed to $PT_4$ patch translator to produce the the resultant feature $X_{F4} \in \mathbb{R}^{H,W,3}$. The $X_{F4}$ contains the super-resolution features $\in \mathbb{R}^{H, W,3}$ scale and is used for generating a super-resolution image by combining this super-resolution feature $X_{F4}$ with input image $X_R$ using the skip-connection. In this work, the generator module $G_{R2S}$ generates super-resolution images at the $2\times$ scale. Therefore this module has been used twice with the same parameters to generate the super-resolution images at the $4\times$ scale.

\subsection{Transformer based Discriminator} \label{Transformer based Discriminator}
The discriminator network in the proposed PTSR is based on the vision transformer \cite{vit}. The vision transformer based discriminator network has been successfully utilized in ViTGAN \cite{vitgan} for image generation task. As the vision transformer was originally proposed by utilizing the patches of the images for image recognition, it is a better suitable network for distinguishing the fake image samples ($Y_S$) from real image samples ($Y_R$). This network is used for training the generator more accurately by utilising the loss parameters through this network. Moreover, using vision transformer as discriminator network also matches the required complexity of the transformer based generator network used in the proposed model. It guides the generator network to generate a realistic $Y_S$ image which is hard to be distinguished from a high-resolution $Y_R$ (ground truth) image. We refer the vision transformer based discriminator network for image super-resolution as $D_T$ network and illustrated in Fig. \ref{fig:modeld}. The modules of $D_T$, includes patch module, class token and positional embedding module, and Transformer Encoder module.


\section{Experiment}
In this section, first we brief the experimental setup and then present the experimental results.
\subsection{Dataset, Metrics and Implements Details}
Following the standard practice, the proposed model is trained on the DIV2K \cite{datadiv} dataset having 800 training images.
The proposed PTSR is evaluated on three benchmark super-resolution datasets, including Set5 \cite{set5}, Set14 \cite{set14}, and BSD100 \cite{bsd100}, which contain 5 images, 14 common used images, and 100 classical test images, respectively. Peak Signal-to-Noise Ratio (PSNR) and Structural Similarity (SSIM) are used to evaluate the performance of the models for the image super-resolution task. These metrics are computed between the generated image $Y_S$ and ground truth image $Y_R$.

We train our model at $2\times$ scale on $X_R \in \mathbb{R}^{256,256,3}$ low-resolution images to $Y_S \in \mathbb{R}^{512,512,3}$ super-resolution images. For $4\times$ scale, first we train on $X_R \in \mathbb{R}^{128,128,3}$ low-resolution images to $Y_S \in \mathbb{R}^{256,256,3}$ super-resolution images and then further use the same trained model at $2\times$ scale to generate the resultant $Y_S \in \mathbb{R}^{512,512,3}$ super-resolution images. In the experiment, we use patch size $k$ as $8\times8$. 
The learning rate is initialised at $2 \times 10^{-4}$ and reduced to 20\% if there is no improvement for 30 epochs.
The Adam optimiser with betas = (0,0.999) is used for training. The bi-linear interpolation with recompute\_scale\_factor at scale 2 is used for up-scaling and down-sampling the patches. Experiments are done using PyTorch with 24Gb NVIDIA GeForce RTX 3090 GPU.

\begin{figure}[!tbh]
\centering
\includegraphics[width=\linewidth,height=150pt]{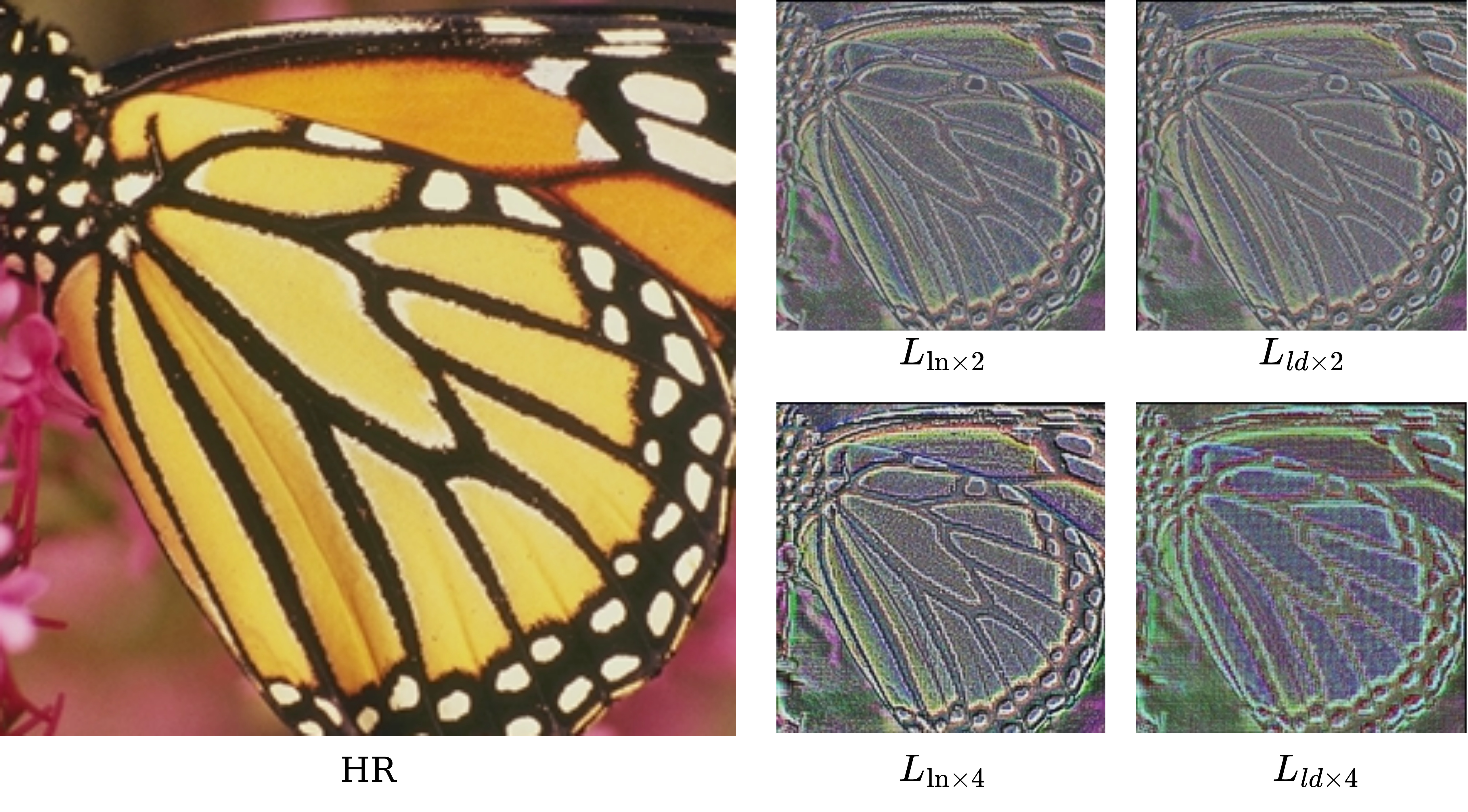}
\caption{Proposed features are further utilised in calculating reconstruction loss for image super-resolution. HR image shows the original ground truth image, $L_{ln\times2}$ and $L_{ln\times4}$ image shows the features that model should learn for $2\times$ and $4\times$ scale, respectively. $L_{ld\times2}$ and $L_{ld\times4}$ image shows the features that model have learned for $2\times$ and $4\times$ scale, respectively} 
\label{fig:loss}
\end{figure}

\begin{figure*}[tbh]
\centering
\includegraphics[width=\linewidth]{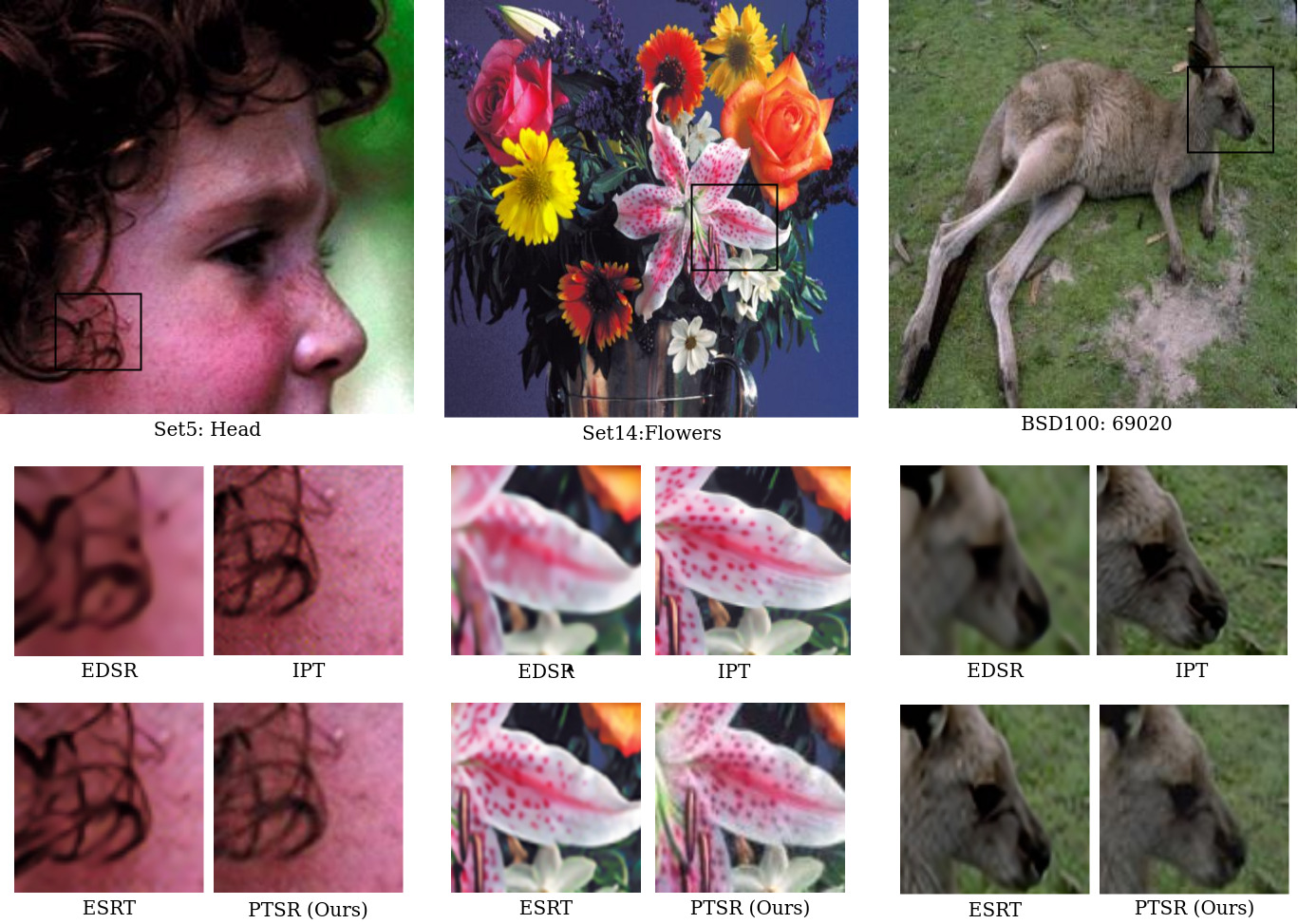}
\caption{Visualization results for $2\times$ super-resolution (a) High-Resolution image, (b) EDSR results, (c) IPT results, (d) ESRT results and (e) PTSR (Ours) results on different dataset images (i) Head image in Set5, (ii) Flowers image in Set14 and (iii) 69020 image in BSD100 dataset}
\label{fig:visual_results}
\end{figure*}

\subsection{Loss Function}To train the PTSR model, we use the Adversarial loss and Reconstruction loss. \paragraph{Generator} The generator loss consists of adversarial generator loss $\mathcal{L}_{G}$  as a binary cross-entropy ($BCE$) loss by classifying the output of discriminator into real category and Reconstruction loss $\mathcal{L}_{R}$ and given as $0.4 \times \mathcal{L}_{G} + 0.6 \times \mathcal{L}_{R}$, where,
\begin{equation}
\mathcal{L}_{G} = BCE(\underset{Y_S \sim \mathbb{P}_g}{\mathbb{E}} [D_T(\uparrow X_R \circled{C} Y_S)],1)
\label{egLG}
\end{equation}
\begin{equation}
\mathcal{L}_{R} = \frac{1}{2H\times2W\times3} \left ( \left\| L_{ln}-L_{ld} \right\|_1 +  \left\| L_{ln}-L_{ld} \right\|_2 \right )
\label{egLR}
\end{equation}
where $\circled{C}$ is channel-wise concatenation, $L_{ln}$ and $L_{ld}$ refer to the features that model should learn and actually learned, respectively. $\mathcal{L}_{R}$ provides better pixel-level generation by measuring the similarity between learning features of the model for $(Y_R$,$Y_S)$ images. A high-resolution image with its learnable feature $L_{ln}$ and learned feature $L_{ld}$ are illustrated in Fig. \ref{fig:loss} for both $2\times$ and $4\times$ super-resolution. 

\paragraph{Discriminator} The adversarial discriminator loss $\mathcal{L}_{D}$ also uses the $BCE$ on the output of discriminator w.r.t. real and generated categories and defined as,
\begin{equation}
\resizebox{\columnwidth}{!}{
$\mathcal{L}_{D} =
\frac{1}{2} (BCE(\underset{Y_S \sim \mathbb{P}_g}{\mathbb{E}} [D_T (\uparrow X_R \circled{C} Y_S],0) + BCE(\underset{Y_R \sim \mathbb{P}_r}{\mathbb{E}} [D_T(\uparrow X_R \circled{C} Y_R)],1)).$
}
\label{egLD}
\end{equation}
This loss has been very effective in generating clear and visually favorable images. It is calculated with an $X_R$ input image condition as in \cite{PIX}.

\subsection{Experimental Results}
\paragraph{Quantitative Evaluation}

In this section, quantitative results of the proposed PTSR method are compared with state-of-the-art methods. Table \ref{tab:PSNR/SSIMx2} shows the PSNR and SSIM using different models on benchmark datasets for $2\times$ super-resolution. The \% improvement using PTSR model over Set5 is 13.317\% in PSNR score \& 1.75\% in SSIM score, over Set14 is 3.25\% in PSNR score \& 0.54\% in SSIM score and BSD100 9.32\% in PSNR score \& 0.21\% in SSIM score for $2\times$ super-resolution as compared to best state-of-the-art method. Table \ref{tab:PSNR/SSIMx2} also shows the PSNR and SSIM using different models for $4\times$ super-resolution. The \% improvement using PTSR model over Set5 is 24.11\% in PSNR \& 6.85\% in SSIM score, over Set14 is 16.44\% in PSNR \& 10.25\% in SSIM score and BSD100 24.44\% in PSNR \& 17.67\% in SSIM score for $4\times$ super-resolution as compared to best state-of-the-art method.

\begin{table}[tbh]
\centering
\caption{PSNR and SSIM comparison among SR method at $2\times$ and $4\times$ scales. Here, * denotes the reproduced results.}
\label{tab:PSNR/SSIMx2}
\resizebox{\columnwidth}{!}{
\begin{tabular}{|c|c|c|c|c|}
\hline
Method & $\times$ & Set5 & Set14 & BSD100\\
& & PSNR/SSIM & PSNR/SSIM & PSNR/SSIM\\
\hline
SRCNN \cite{SRCNNr1} & &  36.66/0.9542 & 32.45/0.9067 & 31.36/0.8879\\
EDSR \cite{EDSR-baseline} & &  37.99/0.9604 & 33.57/0.9175 & 32.16/0.8994\\
CARN \cite{CARN}  & & 37.76/0.9590 & 33.52/0.9166 & 32.09/0.8978\\
ESRT \cite{ESRT} & & 38.03/0.9600 & 33.75/0.9184 & 32.25/0.9001\\
RCAN \cite{RCAN} & 2 & 38.27/0.9614 & 34.12/0.9216 & 32.41/0.9027\\
OISR-RK3 \cite{OISR-RK3} & & 38.21/0.9612 & 33.94/0.9206 & 32.36/0.9019\\
SAN \cite{SAN} & & 38.31/0.962 & 34.07/0.9213 & 32.42/0.9028\\
IGNN \cite{IGNN} & & 38.24/0.9613 & 34.07/0.9217 & 32.41/0.9025\\
Swin-IR \cite{swinir} & & 38.35/0.9620 &  34.14/0.9227 & 32.44/0.9030 \\
IPT \cite{IPT} & &  \underline{38.37}/\underline{0.967$^*$} & \underline{34.43}/\underline{0.924$^*$} & \underline{32.48}/\underline{0.943$^*$}\\
PTSR (Ours) & & \textbf{43.48/0.984} & \textbf{34.55/0.929} & \textbf{35.51/0.945}\\
\hline
SRCNN \cite{SRCNNr1} & & 30.48/0.8628 & 27.5/0.7513 & 26.9/0.7101\\
EDSR \cite{EDSR-baseline} & & 32.09/0.8938 & 28.58/0.7813 & 27.57/0.7357\\
CARN \cite{CARN} & & 32.13/0.8937 & 28.6/0.7806 & 27.58/0.7349\\
IMDN \cite{IMDN} & & 32.21/0.8948 & 28.58/0.7811 & 27.56/0.7353\\
ESRT \cite{ESRT} & & 32.19/0.8947 & 28.69/0.7833 & 27.69/0.7379\\
RCAN \cite{RCAN} & 4 & 32.63/0.9002 & 28.87/0.7889 & 27.77/\underline{0.7436}\\
OISR-RK3 \cite{OISR-RK3} & & 32.53/0.8992 & 28.86/0.7878 & 27.75/0.7428\\
SAN \cite{SAN} & & \underline{32.64}/\underline{0.9003} & 28.92/0.7888 & 27.78/\underline{0.7436}\\
IGNN \cite{IGNN} & & 32.57/0.8998 & 28.85/\underline{0.7891} & 27.77/0.7434\\
Swin-IR \cite{swinir} & & 32.72/0.9021 &  28.94/0.7914 & 27.83/0.7459 \\
SPSR \cite{SPSR} & &  30.400/0.8627 & 26.640/0.7930 & 25.505/0.6576\\
IPT \cite{IPT} & & \underline{32.64}/0.8260$^*$ & \underline{29.01}/0.6783$^*$ & \underline{27.82}/0.6800$^*$\\
PTSR (Ours) & & \textbf{40.510/0.962} & \textbf{33.780/0.870} & \textbf{34.621/0.875}\\
\hline
\end{tabular}}
\end{table}

\begin{figure*}[tbh]
\centering
\includegraphics[width=\linewidth]{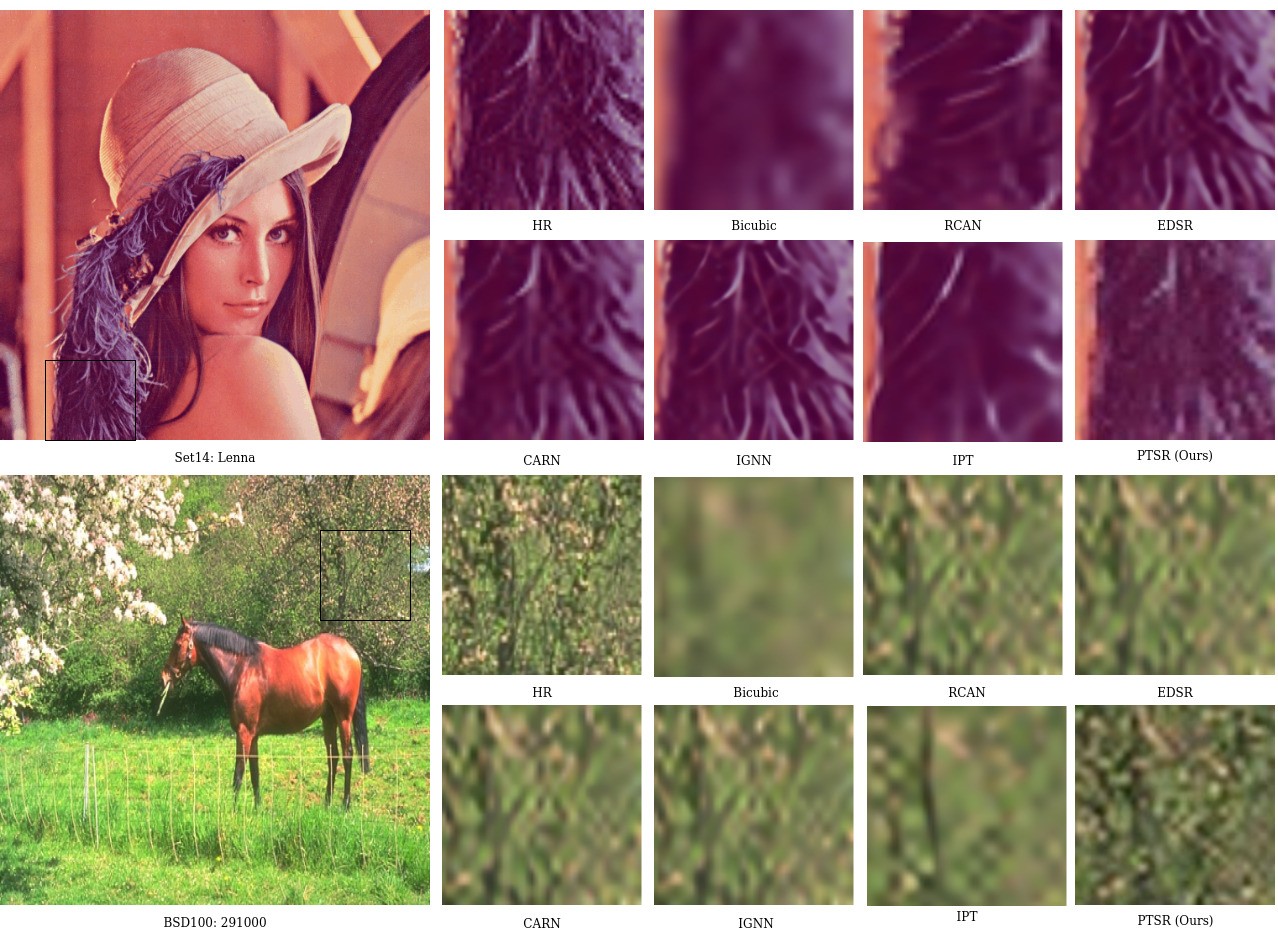}
\caption{ Visualization results for $4\times$ super-resolution. It includes the croped image of HR image, other state-of-the-art results, and PTSR (Ours) results on different dataset images (i) Leena image in Set14 and (ii) 291000 image in BSD100 dataset.}
\label{fig:model4x}
\end{figure*}

It is noticed that the proposed method outperforms the state-of-art models on Set5, Set14, and BSD100 datasets for both $2\times$ and $4\times$ image super-resolution. Though IPT is also a transformer-based model, the proposed PTSR has an edge due to the use of patch-based translator which preserves the local context and at the same utilizes the global context and adversarial training. The existing CNN models are not able to perform very well due to lack of proper global information encoding and adversarial training. However, the existing GAN-based methods exploit the adversarial training, but lacks in terms of global information encoding. The proposed PTSR is able to exploit local context, global information and adversarial training in order to achieve the significant performance. 

The result of different loss function is given in ablation study. Image Super-resolution Reconstruction loss performs best with Adversarial loss for the super-resolution problem with 8.2\% and 2.6\% improvement in terms of in PSNR and SSIM, respectively.


\begin{figure}[tbh]
\centering
\includegraphics[width=\linewidth,height=206pt ]{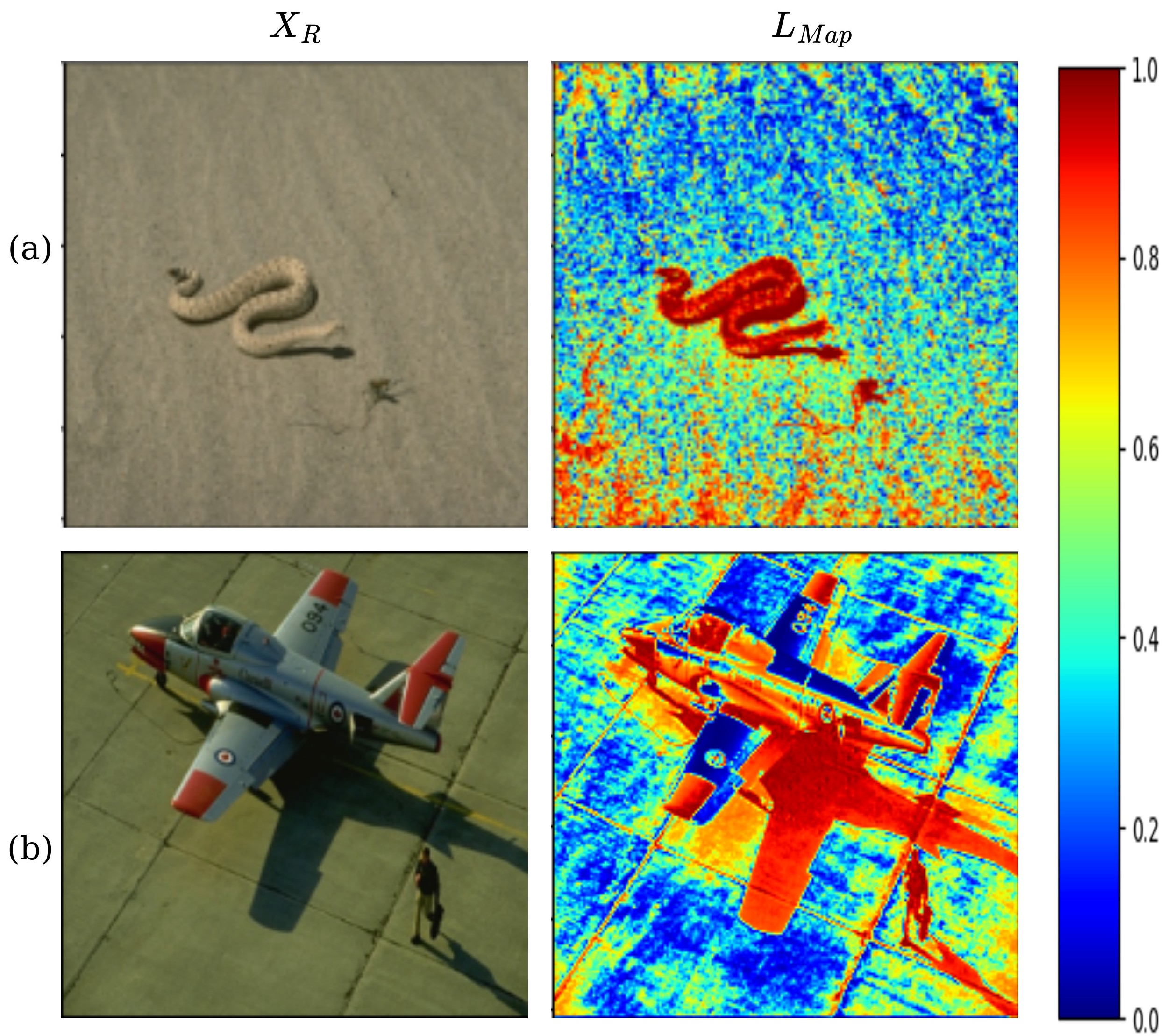}
\caption{Visual Activation Maps for proposed PTSR highlight the regions in input images where the model concentrates more. The blue colour represents the least, and the red colour represents the most important regions in terms of the image super-resolution.}
\label{fig:Lmap}
\end{figure}

\paragraph{Qualitative Evaluation} 
Qualitative results of the proposed PTSR method are compared with state-of-the-art methods, such as EDSR \cite{EDSR-baseline}, ESRT \cite{ESRT}, and IPT \cite{IPT} in Fig. \ref{fig:visual_results} for $2\times$ super-resolution. The proposed method has comparable results with the state-of-the-art in terms of the visual quality in spite of being trained from scratch on a smaller super-resolution dataset (i.e., DIV2K having 800 images). However, IPT \cite{IPT} performs pre-training on large-scale and diverse datasets and is further fine-tuned for the super-resolution. Hence, considering the training complexity and visual quality, the proposed model shows a better trade-off.

The qualitative analysis for $4\times$ super-resolution is shown in the Fig. \ref{fig:model4x}. This figure shows comparison on  different super-resolution dataset such as Set5, Set14 and BSD100. In this we have compared the qualitative result of the proposed PTSR with state-of-the-art such as RCAN \cite{RCAN}, EDSR \cite{EDSR-baseline}, CARN \cite{CARN}, IGNN \cite{IGNN} and IPT \cite{IPT}. The proposed method produces the high visual quality result for $4\times$ super-resolution as compared to state-of-the-art results.

\paragraph{Visual Activation Map}
The $V_{Map}$ visual activation map highlights the regions where the model focus on super-resolution image generation and given as: 
$V_{Map} = \eta ( \nabla (G_{R2S}(X_R)) )\sim (0,1)$,
where $\nabla$ is the cost difference between the $Y_R$ and $Y_S$ for the gradient at $X_R$ image and normalised $\eta$ in the range of (0,1). The $V_{Map}$ is shown in Fig. \ref{fig:Lmap} for $X_R$ image, which shows the model focuses on finer regions for better $Y_S$ image generation.

\subsection{Ablation Study}
\paragraph{Impact of loss function} We conduct different experiments based on the different combination of loss functions for both $2\times$ and $4\times$. The results of different combination of loss function is shown in Table \ref{tab:AblationLoss} (a) for $2\times$ super-resolution. We have tested the results for Adversarial loss $\mathcal{L}_{A}$ and  Image Super-resolution Reconstruction loss $\mathcal{L}_{R}$ and Triplet loss $\mathcal{L}_{T}$. In Table \ref{tab:AblationLoss} (a) $\mathcal{L}_{R1}$ refers to reconstruction loss with $L_1$ regularization and $\mathcal{L}_{R2}$ refers to reconstruction loss with $L_2$ regularization. Where the introduced Image Super-resolution Reconstruction loss performs best with Adversarial loss for the super-resolution problem. Here, 8.2\% in PSNR and 2.6\% in SSIM is improved by the utilising the loss. The image reconstruction loss provides better pixel-level generation by measuring similarity between learning features of the model for super resolution problem.

\paragraph{Impact of Transformer Stack} We conduct different experiments based on the number of transformer Stack for both $2\times$ and $4\times$. In the proposed model we have used '5' stack. The results of different experiment based on of number of transformer stack is shown in Table \ref{tab:AblationLoss} for $2\times$ super-resolution. It shows that among the experiment 5 stack of the transformer performs better in the super resolution domain.

\begin{table}[!t]
\centering
\caption{Impact of different loss function and transformer stack. Comparison is based on PSNR and SSIM. Here, \# Denotes the selected parameter for proposed model.} 


\label{tab:AblationLoss}
\resizebox{\columnwidth}{!}{
\begin{tabular}{|c|c|c|c|}
\hline
\multicolumn{4} {|c|} {(a) Impact of Different Loss Function}\\
\hline
 Loss  & Set5 & Set14 & BSD100 \\
 Function & PSNR/SSIM & PSNR/SSIM & PSNR/SSIM \\
\hline
 $\mathcal{L}_{A}$+$\mathcal{L}_{R1}$ & 39.048/0.965 & 32.357/0.902 & 33.517/0.916  \\
 $\mathcal{L}_{A}$+$\mathcal{L}_{R2}$ & 39.133/0.958 & 32.596/0.901 & 33.783/0.918  \\
 $\mathcal{L}_{A}$+$\mathcal{L}_{R}^\#$ & \textbf{43.48/0.984} & \textbf{34.55/0.929} & \textbf{35.51/0.945} \\
 $\mathcal{L}_{A}$+$\mathcal{L}_{R}$+$\mathcal{L}_{T}$ & \underline{ 41.635/0.979} & \underline{33.524/0.919} & \underline{34.580/0.934} \\
\hline

\multicolumn{4} {|c|} {(b) Impact of Different Transformer Stack}\\

\hline
 Stack  & Set5 & Set14 & BSD100 \\
 & PSNR/SSIM & PSNR/SSIM & PSNR/SSIM \\
\hline
 3 & 42.816/0.983 & 34.142/0.926 & 35.083/0.941 \\
 $5^\#$ & \textbf{43.48/0.984}  & \underline{34.55}/\textbf{0.929} & \underline{35.51}/\textbf{0.945}  \\
 7 & \underline{43.447}/\textbf{0.984} & \textbf{34.616/0.929} & \textbf{35.577}/\underline{0.948}  \\
\hline
\end{tabular}}
\end{table}

\section{Conclusion}
We have proposed a novel patch translator-based GAN architecture (PTSR) for image super-resolution. The PTSR contains two transformer networks, including a generator and a discriminator. The generator contains a patch translator module capable of translating image patches with the help of a transformer. The proposed patch translator-based generator network transforms patches into embeddings, passes through the transformer layer, and converts them back into patches. The proposed patch translator preserves the local context, exploits the global information and enjoys the adversarial training. We have achieved promising results for $2\times$ and $4\times$ resolution. The average \% improvement using PTSR model for $2\times$ super-resolution by 8.62\% in PNSR \& 0.83\% in SSIM score and for $4\times$ super-resolution by 21.66\% in PNSR \& 11.59\% in SSIM score, as compared to the best competitive models. It is noticed based on saliency map that the proposed model produces finer details in the super-resolution images. 
This model can be  used for various super-resolution applications. 
Also, the patch translator module can be used to propose other transformer-based image-to-image translation tasks.

\section*{Acknowledgement}
The authors would like to thank Ministry of Education, Govt. of India for providing the financial support to carry out this research at Indian Institute of Information Technology, Allahabad.

\bibliographystyle{IEEEtran}
\bibliography{References}

\begin{IEEEbiography}[{\includegraphics[width=1in,height=1.25in,clip,keepaspectratio]{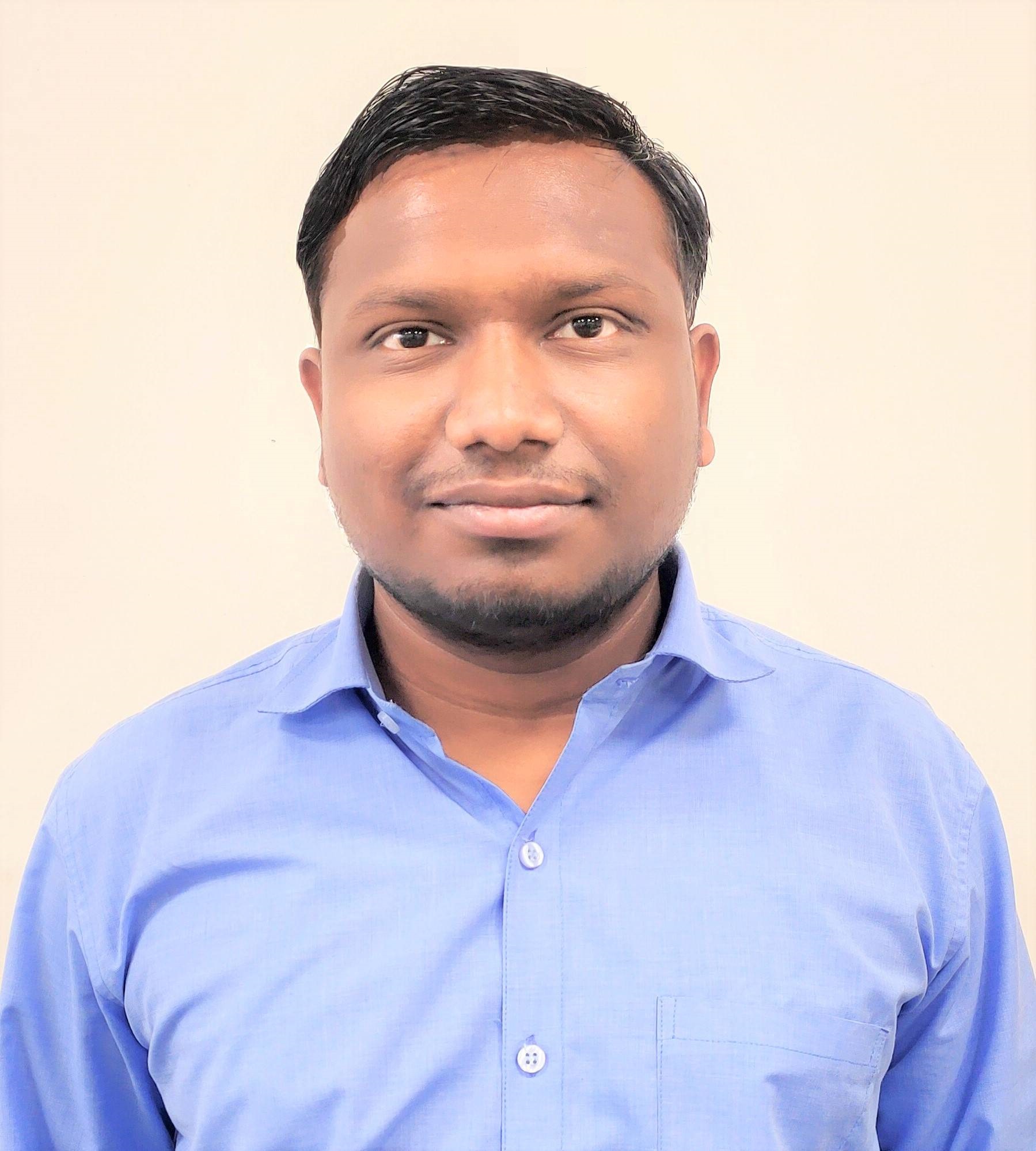}}]{Neeraj Baghel} is currently associated with Computer Vision and Biometrics Lab, Indian Institute of Information Technology Allahabad as Research Scholar. Earlier, he has served as JRF at IIIT SriCity over the DRDO Young Scientist Laboratory. He has also served as JRF at Centre for Advanced Studies, Lucknow. Currently working in the areas of Artificial Intelligence, Computer Vision, Image \& Video Processing, Super-Resolution and their applications. Neeraj is currently offering his volunteer service as Chair at IEEE Student Branch IIIT Allahabad, SAC member at IEEE UP Section.
\end{IEEEbiography}

\begin{IEEEbiography}[{\includegraphics[width=1in,height=1.25in,clip,keepaspectratio]{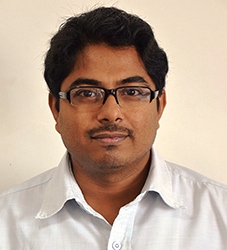}}]{Shiv Ram Dubey} is with the Indian Institute of Information Technology (IIIT), Allahabad since July 2021, where he is currently the Assistant Professor of Information Technology. He was with IIIT Sri City as Assistant Professor from Dec 2016 to July 2021 and Research Scientist from June 2016 to Dec 2016. He received the PhD degree from IIIT Allahabad in 2016. Before that, from 2012 to 2013, he was a Project Officer at Indian Institute of Technology (IIT), Madras. He was a recipient of several awards including the Best PhD Award in PhD Symposium at IEEE-CICT2017, Early Career Research Award from SERB, Govt. Of India and NVIDIA GPU Grant Award Twice from NVIDIA. Dr. Dubey is serving as the Treasurer of IEEE Signal Processing Society Uttar Pradesh Chapter.
His research interest includes Computer Vision and Deep Learning.
\end{IEEEbiography}

\begin{IEEEbiography}[{\includegraphics[width=1in,height=1.25in,clip,keepaspectratio]{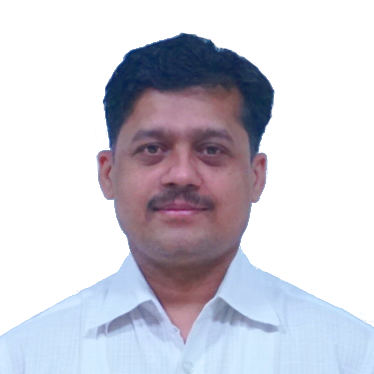}}]{Satish Kumar Singh}
is with the Indian Institute of Information Technology Allahabad, as an Associate Professor from 2013 and heading the Computer Vision and Biometrics Lab (CVBL). Earlier, he served at Jaypee University of Engineering and Technology Guna, India from 2005 to 2012. His areas of interest include Image Processing, Computer Vision, Biometrics, Deep Learning, and Pattern Recognition. Dr. Singh is proactively offering his volunteer services to IEEE for the last many years in various capacities. He is the senior member of IEEE. Presently Dr. Singh is the Section Chair IEEE Uttar Pradesh Section (2021-2022) and a member of IEEE India Council (2021). He also served as the Vice-Chair, Operations, Outreach and Strategic Planning of IEEE India Council (2020) \& Vice-Chair IEEE Uttar Pradesh Section (2019 \& 2020). Prior to that Dr. Singh was Secretary, IEEE UP Section (2017 \& 2018), Treasurer, IEEE UP Section (2016 \& 2017), Joint Secretary, IEEE UP Section (2015), Convener Web and Newsletters Committee (2014 \& 2015). 
Dr. Singh is also the technical committee affiliate of IEEE SPS IVMSP and MMSP and presently the Chair of IEEE Signal Processing Society Chapter of Uttar Pradesh Section. 
\end{IEEEbiography}

\end{document}